\documentclass[journal,12pt,onecolumn,draftclsnofoot,]{IEEEtran}

\usepackage{cite,url}
\usepackage{amssymb,amsmath,amsfonts,amssymb,verbatim,graphicx,bm,color,amsbsy,amsthm,algorithm,algorithmic,epsfig,subfig}
\usepackage{epstopdf}
\usepackage{caption}
\usepackage[inline]{enumitem}
\usepackage{empheq}
\usepackage{subfiles}
\usepackage{booktabs}
\newcommand{\ba}{{{\bf a}}}

\newtheorem{theorem}{Theorem}
\newcommand*\widefbox[1]{\fbox{\hspace{1em}#1\hspace{1em}}}

\begin{document}

\title{Phase noise in modular millimeter wave massive MIMO}
	

\author{\IEEEauthorblockN{Maryam Eslami Rasekh\IEEEauthorrefmark{1}, Mohammed Abdelghany\IEEEauthorrefmark{2}, Upamanyu Madhow\IEEEauthorrefmark{3}, Mark Rodwell \IEEEauthorrefmark{4}}\\
\IEEEauthorblockA{Department of Electrical and Computer Engineering\\
	University of California Santa Barbara\\
	Email: \{\IEEEauthorrefmark{1}rasekh, \IEEEauthorrefmark{2}mabdelghany, \IEEEauthorrefmark{3}madhow, \IEEEauthorrefmark{4}rodwell\}@ucsb.edu}
}

\maketitle
\begin{abstract}
		 This paper investigates the effect of oscillator phase noise on a multiuser millimeter wave (mmWave) massive MIMO uplink as we scale up the number of base station antennas, fixing the load factor, defined as the ratio of the number of simultaneous users to the number of base station antennas.  We consider a modular approach in which the base station employs an array of subarrays, or ``tiles.'' Each tile supports a fixed number of antennas, and can therefore be implemented using a separate radio frequency integrated circuit (RFIC), with synchronization across tiles accomplished by employing a phased locked loop in each tile to synthesize an on-chip oscillator at the carrier frequency by locking on to a common lower frequency reference clock. Assuming linear minimum mean squared error (LMMSE) multiuser detection, we provide an analytical framework that can be used to specify the required power spectral density (PSD) mask for phase noise for a target system performance. Our analysis for the phase noise at the output of the LMMSE receiver indicates two distinct effects: self-noise for each user which is inversely proportional to the number of tiles, and cross-talk between users which is insensitive to the number of tiles, and is proportional to the load factor.  These analytical predictions, verified by simulations for a $\bf{140}$ GHz system  targeting a per-user data rate of $\bf{10}$ Gbps, show that tiling is a robust approach for scaling. Numerical results for our proposed design approach yield relatively relaxed specifications for phase noise PSD masks.

Keywords: Phase noise, millimeter wave, THz, multiuser, massive MIMO, 5G, next generation wireless, modular, array of subarrays.
\end{abstract}

\section{Introduction} \label{sec:intro}

\begin{figure}
	\centering
	\includegraphics[width=.4\columnwidth]{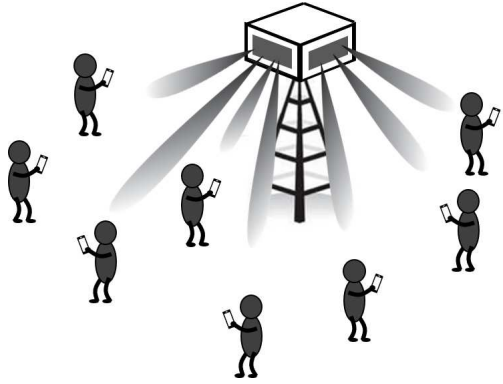}
	\caption{Mmwave multiuser massive MIMO.}
	\label{fig:picocell}
\end{figure}

The emergence of millimeter wave communication has produced unprecedented possibilities for next generation mobile networks. 
In addition to the large amounts of available spectrum, much of it unlicensed, the band has immense potential for spatial multiplexing.
The small wavelengths ($5$ mm at 60 GHz, only $2$ mm at $140$ GHz) imply that hundreds or even thousands of antenna elements can fit on relatively 
small platforms, producing massive electronically steerable arrays with very small beamwidth.  Most prior research in mmWave systems assumes
RF beamforming, which employs one RF chain for the entire array, or hybrid beamforming, which employs a number of RF chains which is much smaller
than the number of antenna elements in the array. However, advances in silicon implementations of mmWave hardware imply that, at least for a moderate number of antenna elements, it is possible to build low-cost RFICs with one RF chain for each antenna, opening up the possibility of all-digital beamforming for multiuser MIMO. 
In this paper, we investigate modular architectures using such RFICs as ``tiles,'' in a regime where the number of antennas per tile is fixed, but the size
of the overall antenna array is scaled up by increasing the number of tiles.  We consider a mmWave massive MIMO uplink, shown in Figure \ref{fig:picocell},
in which the number of simultaneous users scales with the antenna array size, and our goal is to understand whether phase noise is a bottleneck for scaling.

While additive noise can be averaged out, and multiuser interference suppressed, using degrees of freedom, the multiplicative nature of phase noise leads to distortion that scales with signal power, and hence to performance floors that can only be alleviated by reducing oscillator noise.  Since realizing low-noise oscillators
is challenging at higher carrier frequencies, we are interested in how much we can relax phase noise
specifications while attaining a target system-level performance. To this end, we develop in this paper a framework for analyzing the impact of phase noise in massive MIMO, modeling its propagation in the proposed tiled architecture. Each tile is controlled by a separate RF chip that performs down conversion as well as analog-to-digital conversion, alleviating the need to transport analog RF signals across the entire frontend. In order to emulate a single large array, the tiles must be synchronized in frequency and phase, meaning they need to be locked to a common reference. We are interested in base station arrays with hundreds of antennas (e.g., horizontal scanning with half-wavelength spacing between elements), and distributing a stable clock at mmWave carrier frequencies across the array is challenging and power-inefficient. Instead, a low-frequency reference is distributed to the tiles, and is multiplied up to the carrier frequency via a PLL on each tile. 
Thus, the sources of phase noise in this architecture are from the common low-frequency reference, and from the VCOs driving the PLLs in the tiles.


\noindent 
{\it Concept system:} While our analytical framework is general, our numerical evaluations are based on a concept system operating at a carrier frequency of 140 GHz, with common low-frequency reference at 10 GHz.  We consider single carrier QPSK modulation at 5 Gbaud symbol rate, corresponding to an uncoded bit rate of 10 Gbps per user, 
and $N=256$ base station antennas per sector.  The load factor $\beta$, defined as the ratio of number of simultaneous users 
to the number of antennas, varies from $\beta = \frac{1}{16}$ to $\beta = \frac{1}{2}$ (with a nominal value $\beta = \frac{1}{4}$), which corresponds to sector-level uncoded throughputs ranging from 160 Gbps to 1.28 Tbps.  These aggressive specifications may well be beyond what is required in deployed systems, but they allow us to explore
the limits of system performance with reasonable hardware requirements.  For the load factors of interest, spatial matched filtering does not yield acceptable performance, hence we consider LMMSE reception, arguably the simplest multiuser detection strategy we could adopt in exploring these limits.  

We now summarize our contributions, and then place these in the context of prior literature.

\subsection{Contributions}

\begin{figure}
	\centering
	\includegraphics[width=0.55\columnwidth,trim={1in -0.in 1.in 0.in},clip]{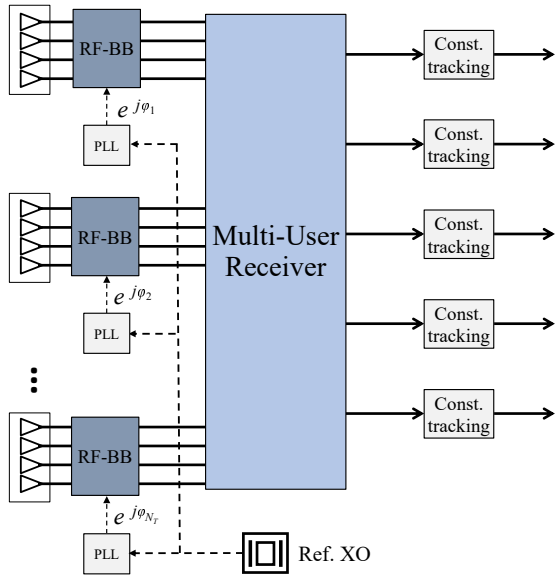}
	\caption{Architecture of the tiled multiuser massive MIMO receiver.}
	\label{fig:architecture}
\end{figure}

We propose the architecture depicted in Figure \ref{fig:architecture}, and model the propagation of phase noise through each of the blocks
depicted therein: the tile-level PLLs, the LMMSE
multiuser detector, and decision-directed phase drift tracking for each user's stream at the LMMSE output.  
We provide simple yet accurate estimates of its impact on system performance.  The resulting analytical framework is used to provide design guidelines that greatly simplify the complicated task of joint hardware/system development.
This includes determining maximum phase noise PSD masks for oscillators that guarantee the desired performance level for a given system configuration.

Our focus is on the impact of phase noise on demodulation, and we assume that the variations of the spatial channels for the users
are much slower than those of the phase noise contributions at the receiver.  Thus, we assume 
ideal channel estimates at the base station (which might, for example, be obtained at the beginning of a frame) which enables us to compute
a ``naive'' LMMSE receiver which does not account for, or track, subsequent variations due to phase noise.  We also consider an LMMSE receiver which
accounts for the statistics of the phase noise, which is what would be learnt by a continuously adaptive implementation. 

In the absence of phase noise, LMMSE reception suppresses multiuser interference (and forces it to zero at high SNR), and avoids high SNR performance floors.
Phase noise, on the other hand, leads to impairments which
scale with signal strength, leading to performance floors.  
However, our analysis shows that this performance floor can be mitigated by increasing the number of tiles
and lowering the load factor. Indeed, a key conclusion from the scaling laws we derive is that, for a given oscillator quality,
phase noise is {\it less} harmful for massive MIMO than for a single-input single-output (SISO) system, and does not represent
a bottleneck for our ambitious system goals.  

The technical approach and results that lead us to these conclusions are summarized as follows:\\
$\bullet$ 
Using a standard linearized model for the PLL in each tile, we show that the PLL acts as a lowpass filter for the reference phase noise, and as a highpass filter for the VCO phase noise.\\
$\bullet$ The contribution of the reference phase noise at the PLL output passes unchanged through the LMMSE multiuser detector, and is highpass filtered
by the constellation tracking block.  Given the lowpass filtering due to the PLLs, the impact of reference phase noise on system performance becomes negligible.
Performance is therefore limited by the VCO phase noises at the tile PLL outputs, which is highpass.\\
$\bullet$ The performance of ``naive'' LMMSE and LMMSE accounting for phase noise statistics are virtually identical, which implies that 
continuously adaptive implementations are not required.\\
$\bullet$ Under a small phase approximation for the VCO phase noise (accurate for the design regimes of interest), we compactly
characterize the LMMSE output of user $k$ as follows (ignoring additive noise in this summary):
$$
y_k = e^{j{\psi}_k} s_k + {I}_k
$$
where the phase rotation $\psi_k$ is self-noise from user $k$ with variance  $\frac{\sigma_\phi^2}{N_T}$, and 
the additive interference $I_k$ is due to other users, with variance $\beta\sigma_\phi^2$,
with $\sigma_{\phi}^2$ denoting the VCO phase noise variance at the PLL output.
Thus, the self-noise is reduced by increasing the number of tiles, while multiuser interference is insensitive to the number of tiles,
but is reduced by decreasing the load factor.\\
$\bullet$ We provide an upper bound on the error probability for QPSK in a SISO system, in terms of an effective SNR which combines the effect of
additive noise and phase noise.  We then use this to obtain an accurate approximation to the error probability for our massive MIMO model, using
an effective SINR combining phase noise and additive noise, and including the effect of noise enhancement due to interference suppression.\\
$\bullet$ We provide design examples on how to determine the maximum tolerable VCO phase noise variance $\sigma_{\phi}^2$ for a desired
system-level performance, and how that maps to specifications on the allowable phase noise PSD. Numerical results show that the performance
floor for an uncoded system is well below $10^{-3}$ for the regimes of interest, which can be easily handled by a lightweight high-rate error correction code.

\subsection{Related Work}
The effect of various hardware impairments (such as amplifier nonlinearity, low precision ADC and DAC, I-Q imbalance, and phase noise) on massive MIMO systems has been the subject of many studies, including \cite{abdelghany2018towards,jacobsson2017throughput,bjornson2014massive,moghadam2018energy,wu2016hardware} to name a few. Among these effects, phase noise is particularly challenging due to its multiplicative nature, especially for large communication bandwidth.
Significant efforts have been made by the hardware community to extract accurate models for phase noise from the physics of oscillator circuitry. Various works have utilized the framework established in \cite{ leeson1966simple} to develop descriptions for the phase noise generated in different configurations \cite{razavi1996study,hajimiri1998general, mehrotra2000noise, kim2008phase, herzel2010analytical, kalia2011simple}, typically by describing the PSD of the phase noise process. Modeling the overall impact of phase noise on the communication link requires a system-level analysis that incorporates such models into the signal reception and decoding process.
Prior studies in this direction, however, have often employed oversimplified models such as the pessimistic Wiener process model considered in \cite{bjornson2015massive} or the white noise model assumed in \cite{thomas2010phase}. 
Many studies neglect the possibility of leveraging time domain correlations in a phase noise process for suppressing phase drift, unnecessarily tying system performance to channel tracking overhead by relying on frequent CSI updates for drift suppression \cite{ combes2017approximate}, but  
tracking phase drift over time is crucial in realizing the true capacity of a practical system \cite{wang2017zf}.
Furthermore, most studies that consider multiple antenna systems assume either a common clock with fully correlated phase noise across the array (synchronous clock distribution), or a free-running oscillator at each element untethered to a common reference (asynchronous distribution), producing uncorrelated phase drifts at different antennas \cite{wang2016simple,wang2017zf , pitarokoilis2014uplink,thomas2010phase, krishnan2016linear }. Synchronous clocking is impractical at high carrier frequencies while the asynchronous configuration suffers from beamforming degradation as a result of rapid ``channel aging". 

In terms of modeling, the closest approach to ours in the literature is that of \cite{puglielli2016phase}, which investigates the effect of phase noise on 
OFDM multi-user beamforming arrays. They assume an independent oscillator at each array element locked to a common lower frequency reference via a PLL multiplier and model the filtering effects of the PLL on common and independent phase noise.  
In their analysis, the authors rely on a subset of subcarriers acting as pilots for phase noise tracking, and derive SINR predictions for single user and multiuser beamforming that predict similar scaling laws as our analysis. However, as shown in this and other studies, phase noise poses an inherent challenge for OFDM systems: The inter-carrier interference caused by phase noise increases with the number of subcarriers posing a fundamental limit on the bandwidth of an OFDM system impacted by phase noise  \cite{schenk2005influence, puglielli2016phase, petrovic2007effects, wu2004ofdm, robertson1995analysis, madani2010analysis, sathananthan2001performance, tomba1998effect, pollet1995ber, armada1998phase }.  OFDM also has other drawbacks for mmWave systems: the linearity required to handle high peak-to-average ratios is difficult to realize at reasonable power efficiency at high frequencies, and the precision required in analog-to-digital conversion is a challenge at large
bandwidths.  We therefore focus on a single carrier system in this paper.

A key distinguishing aspect of the approach in the present paper is that it abstracts the oscillator phase noise models developed by hardware experts into a rigorous system-level framework in which we model the propagation of phase noise through signal processing blocks.  To the best of our knowledge, this is also the
first paper to consider a hierarchical approach to scaling massive MIMO with a fixed number of antennas per tile. The analysis is distilled into compact scaling laws for modular MIMO which clearly link performance to hardware and system design parameters.

The current work is a significant extension of preliminary results reported in our conference paper \cite{rasekh2019phase}, including a more detailed theoretical treatment of phase noise scaling, introduction of BER approximations via effective SNR and SINR, derivation of LMMSE reception accounting for phase
noise statistics, and validation of our analytical estimates via full system simulations.

\section{System Model} \label{sec:system_model}
We consider uplink multiuser MIMO in which the base station, equipped with an $N$-element digitally steered array, 
simultaneously receives signals sent by $K$ users. 
The \textit{load factor} is defined as the ratio of users to array size, and denoted by $\beta=K/N$.

We consider line-of-sight (LoS) channels between users and the base station. We focus on the impact of phase noise on multiuser demodulation rather 
than on channel estimation.  Thus, we consider durations over which the spatial channels are well modeled as constant, and assume that the spatial channels at the beginning of such durations are known to the receiver.   We discuss this assumption further in our conclusions in Section \ref{sec:conclusions}.
The vector $\mathbf{h}_k$ represents the channel of user $k$, with complex amplitude $\alpha_k$ and angle of arrival $\theta_k$. For ease of notation we define the \textit{spatial frequency} corresponding to $\theta_k$ as $\omega_k = (2\pi d/\lambda) \sin\theta_k$, where $d$ is the array inter-element spacing and $\lambda$ is the carrier wavelength. User $k$'s channel can thus be represented by the sinusoid,
\[
\mathbf{h}_{k} = g_k \left[\, 1, e^{j\omega_k}, \, e^{j 2 \omega_k },\, \dots,\, e^{j (N-1) \omega_k}\, \right]^T.
\]
For simplicity, we assume that the $K$ spatial frequencies are distributed uniformly over $(-\pi,\pi)$ (rather than uniformly over angles of arrival).
In order to limit the variation in performance across users, we assume that users that are too close in spatial frequency are orthogonalized in time or frequency domain. and maintain a minimum pairwise distance of $2\pi/N$ in spatial frequency between users. While our analytical framework easily accommodates variations in user power,
we assume perfect power control for simplicity of exposition.

For the concept system described in Section \ref{sec:intro} our nominal configuration is $K=64$, or $\beta=1/4$.
We consider single-carrier digital modulation with Gray coded QPSK, and use uncoded BER of $10^{-3}$ as our performance target, since low frame error
rates can be achieved using high-rate error correction codes at this BER. We assume here that the bandwidth $B$ (5 GHz for our concept system) equals the symbol rate $1/T_\text{symb}$, but the analysis easily extends to accommodate excess bandwidth.
Nominal beamformed SNR (not including phase noise) is $14$ dB. 
This provides a margin of $4$ dB compared to the $10$ dB required SNR for $10^{-3}$ BER for a SISO link without multiuser interference or phase noise.
LMMSE reception is used to separate user data streams, and constellation tracking with window size of $10$ is performed at the output of each channel to offset slow-time-scale oscillator phase drift.

While we use uncoded BER as our metric here, our SINR-based analytical framework easily extends to alternative metrics such as spectral efficiency.





\textbf{The modular architecture.}
Figure \ref{fig:architecture} depicts the modular structure of an $N$-element array 
containing $N_T$ tiles, each with $N_0$ elements, so that
$N=N_TN_0$. Our nominal configuration is $N_T=16$ tiles and $N_0=16$ elements per tile. We define the ``{underloaded}" case where the number of users is no larger than the tile size, i.e., $K\le N_0$ or equivalently $\beta N_T\le 1$. Since our goal is to scale up the system ($K,N\rightarrow\infty$) while keeping tile size $N_0$ and load factor $\beta$ constant, 
the underloaded regime will not be the operating condition of a large system and is of limited interest for scaling.

A $10$ GHz reference is distributed to tiles and frequency multiplied on-tile using a PLL-controlled VCO to produce the $140$ GHz carrier.
Since the phase noises at different VCOs are independent, the carriers at different tiles contain independent phase noise components. 
The system block diagram for this process is depicted in Figure \ref{fig:PLL}.
In our running example, the multiplication factor is $N_f = 14$, producing a $140$ GHz carrier from the $10$ GHz reference clock.  
The PLL is type-2 with loop resonance frequency of $\omega_n=1$ MHz and damping factor of $\xi =$ 0.707, achieved by setting 
\[
k_V = N_f \omega_n^2 , \qquad H_\text{LP}(s) = \frac{1-\frac{2\xi}{\omega_n}s}{s}.
\]

\textbf{Receiver modeling.} In the absence of phase noise, the complex baseband signal received on the $N$-dimensional array 
is described by
\begin{equation}\mathbf{x} = \mathbf{H}\mathbf{s} + \boldsymbol{\nu},
	\label{eq:rec_sig}
\end{equation}
where $\mathbf{H}=[\mathbf{h}_1 ...\mathbf{h}_K]$ is the $N\times K$ channel matrix, $\mathbf{s}$ is a $K$-dimensional vector containing the symbols transmitted by users, and
$\boldsymbol{\nu}\sim \mathcal{CN}(\mathbf{0},\sigma_\nu^2 \, \mathbf{I}_N)$ is the additive receiver noise vector. 
A linear receiver, $\boldsymbol\Gamma_{K\times N}$, is used to estimate the transmitted symbols as
\begin{equation} {\mathbf{y}} = \boldsymbol\Gamma \, \mathbf{x}. \label{eq:linear_reception} \end{equation}

In the presence of phase noise, the received signals on separate subarrays are distorted by different phase noise terms during down conversion. Denoting by $\mathbf{H}_i$ the $N_0\times K$ channel matrix of the $i$'th subarray, we have
\[
\mathbf{H} = \left[ 
\begin{array}{c}
\mathbf{H}_1\\
\mathbf{H}_2\\
\vdots \\
\mathbf{H}_{N_T}\\
\end{array}
\right]
\]
and can model phase distorted reception by
\begin{equation}
	{\mathbf{y}} = \boldsymbol{\Gamma}( \left[
	\begin{array}{c}
		\mathbf{H}_1e^{j\phi_1}\\
		\mathbf{H}_2e^{j\phi_2}\\
		\vdots \\
		\mathbf{H}_{N_T}e^{j\phi_{N_T}}\\
	\end{array} \right]
	\mathbf{s} + \boldsymbol{\nu}).
	\label{eq:reception_in_phase_noise}
\end{equation}
Since the additive noise, $\boldsymbol{\nu}$, is a vector of i.i.d. symmetric complex Gaussians, its distribution is not affected by phase noise, therefore we use the same symbol to represent the phase distorted version.

Assuming phase noise terms are small (which we ensure by design for the regimes of interest), we use the approximation $e^{j\epsilon}\approx 1+j\epsilon$ to write (\ref{eq:reception_in_phase_noise}) as
\begin{equation}
	{\mathbf{y}} \approx \boldsymbol{\Gamma}(\mathbf{H}\mathbf{s} + \boldsymbol{\nu} +   \left[
	\begin{array}{c}
		\mathbf{H}_1{j\phi_1}\\
		\mathbf{H}_2{j\phi_2}\\
		\vdots \\
		\mathbf{H}_{N_T}{j\phi_{N_T}}\\
	\end{array} \right]
	\mathbf{s}) .
	\label{eq:reception_in_phase_noise_small_phase}
\end{equation}
Thus we see that phase noise introduces an additional distortion term to the classical multiuser reception model described by (\ref{eq:rec_sig}) and (\ref{eq:linear_reception}), which we model in detail in later sections.

\section{Phase Noise Modeling\label{sec:phase_noise}}


A perfectly noiseless carrier has constant complex baseband amplitude, $C(t) = A$. 
In a noisy setting, the complex envelope of an oscillator output becomes 
$C(t) = A + n(t) = A + n_c(t) + jn_s(t)$,  
where $n(t)$ is a complex Gaussian random process and $n_c$ and $n_s$ are its real and imaginary parts.
For $A \gg |n (t)|$, we obtain our standard phase noise model
\begin{equation*} 
	C(t) = Ae^{j\phi(t)} 
\end{equation*}
where $\phi (t) = \frac{n_s(t)}{A}$ is a Gaussian random process with power expressed in dBc, or dB relative to carrier power.

As is conventional in the hardware literature, we denote the PSD of $\phi$ as $L(f)$.
The dynamics of active components in oscillators  produce \textit{colored} phase noise in the output sinusoid \cite{leeson1966simple,razavi1996study,hajimiri1998general}. This phase noise is modeled as a combination of white noise and lowpass components with PSD proportional to $1/f$, $1/f^2$, and $1/f^3$. Oscillator phase noise PSD is thus described parametrically by
\begin{equation}
	L(f) = a_0 + \frac{a_1}{f} + \frac{a_2}{f^2} + \frac{a_3}{f^3}.
	\label{eq:phase_noise_model}
\end{equation}

For simplicity we assume here that all clocks are unit amplitude ($A=1$).
When a noisy oscillator is used to down-convert an RF signal, the bandpass phase noise in the oscillator output is directly transferred to the demodulated baseband signal. For digital communication this translates to rotation of the baseband symbols relative to the transmitted constellation points.

\subsection{Phase noise in the tiled array}
\begin{figure}
	\centering
	\includegraphics[width=0.55\columnwidth]{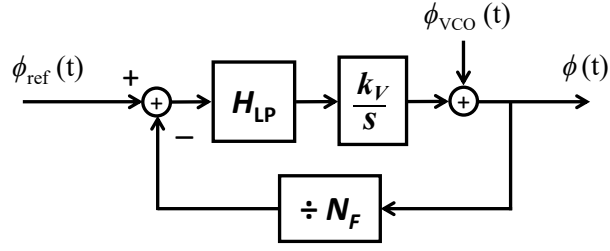}
	\caption{LTI system model of PLL for phase noise.}
	\label{fig:PLL}
\end{figure}

Fig. \ref{fig:PLL} shows the linearized PLL model. The signal described by this model is the \textit{phase} of the input and output signals and therefore predicts accurately how VCO and reference phase noise are affected in the process. The phase noise at the output of this system is the sum of contributions from the reference phase and VCO phase noise; the former is identical in all tiles, whereas the latter is independent from one tile to another but identical over elements of the same tile.

The relation between phase noise at the output of the PLL and reference and VCO phase noise is described through the model of Fig. \ref{fig:PLL} as
\begin{align*}
	& \tilde\phi(s) = \tilde\phi_\text{vco}(s) + \frac{\alpha}{s} H_\text{LP}(s) \left( \tilde\phi_\text{ref}(s) - \frac{\tilde\phi(s)}{N_f}  \right)
\end{align*}
where $\tilde{u}$ denotes the frequency domain representation (Laplace transform) of function $u$.
With some manipulation, we arrive at the PLL filters applied to each phase noise source,
\[
\tilde\phi(s) = H^\text{PLL}_\text{ref}(s) \tilde\phi_\text{ref}(s) + H^\text{PLL}_\text{vco}(s) \tilde\phi_\text{vco}(s) \nonumber \]
where
\begin{align}
	& H^\text{PLL}_\text{ref}(s) = \frac{N_f\alpha H_\text{LP}(s)}{N_fs+\alpha H_\text{LP}(s)} \nonumber \\
	& H^\text{PLL}_\text{vco}(s) = \frac{N_f s}{N_fs+\alpha H_\text{LP}(s)}. \nonumber 
\end{align}
The loop filter, $H_\text{LP}$, is low pass, therefore we observe that the PLL acts as a low pass filter for reference phase noise and a high pass filter for VCO phase noise. 

The filtering of {reference phase noise} by the PLL leaves only its \textit{low frequency} components which results in a slow-varying signal that changes on a time scale of many symbols (hundreds or even thousands, depending on the filter bandwidth and symbol rate). Furthermore, this noise is constant over the array and passes through the linear receiver to the output where it can be tracked and compensated as described in the next section. We see, therefore, that the overall impact of reference phase noise on demodulation is very small.  Of course, as discussed in Section \ref{sec:conclusions}, accounting for reference phase noise is important 
for channel estimation.

{VCO phase noise,} on the other hand, is constant for elements on one tile, but \textit{independent over different tiles,} and therefore affects multiuser detection in a nontrivial manner.
We define by $\phi_i(t)$ the VCO phase noise in the carrier of tile $i$ and derive its variance as
\begin{equation}
	\mathbb{E}\phi_i^2 = \sigma_\phi^2 = \int_{-B/2}^{B/2}L_\text{vco}(f)\left|H^\text{PLL}_\text{vco}(f)\right|^2df
	\label{eq:sigma_phi_2}
\end{equation}
where $L_\text{vco}(f)$ is the VCO phase noise PSD and $B$ is the system bandwidth. 
As we demonstrate in upcoming discussions, the impact of phase noise on the tiled multiuser system is determined by this variance, which is a linear function of the oscillator phase noise coefficients introduced in (\ref{eq:phase_noise_model}),
\begin{align}
	\sigma_\phi^2 &= \sum_{i=0}^{3} q_i a_i,  \label{eq:ph_component_linear}\\
	q_i&=\int_{-B/2}^{B/2}\frac{1}{f^i}\left|H^\text{PLL}_\text{vco}(f)\right|^2 df . \nonumber
\end{align}
After quantifying the largest tolerable phase noise variance, $\sigma_{\phi\,\text{max}}^2$, any $L(f)$ mask that satisfies $\sum q_i a_i\le \sigma_{\phi\,\text{max}}^2$ maintains the desired system performance. The values of these coefficients are given for our nominal configuration in Table \ref{tab:q_coeffs}.

\begin{table} 
	\centering
	\caption{Examples of $q$ coefficients for nominal system.}	
	\begin{tabular}{lcccc}
		\hline \noalign{\smallskip}
		& $q_0$ (Hz) & $q_1$ & $q_2$ (Hz$^{-1}$) & $q_3$ (Hz$^{-2}$) \\
		\noalign{\smallskip}\hline \hline\noalign{\smallskip}
		$ H_\text{vco}^\text{PLL}$ & $5\times10^9$ & $15$ & $2.2\times10^{-6}$ & $1.6\times10^{-12}$ \\
		\noalign{\smallskip} \hline \noalign{\smallskip}
		$ H_\text{ref}^\text{PLL} H_W\,{\color{white}^*} $ & $1.2\times10^7$ & $0.22$ & $6.2\times10^{-8}$ & $1.1\times10^{-13}$ \\ 
		\noalign{\smallskip}
		\hline
		\noalign{\smallskip}
	\end{tabular}\\
	$\qquad \qquad \qquad \sigma_\phi^2 = 1.3\times10^{-1} \qquad \sigma_0^2 = 4.4\times10^{-5}$
	\label{tab:q_coeffs}
\end{table}

We now provide a simple upper bound (Theorem \ref {lemma:equivalent_noise}) on the BER for Gray coded QPSK in a SISO system, defining an equivalent SNR that combines the effects of  phase noise and additive noise. We use this result in Section \ref{sec:interference}, where we characterize the SINR in our MIMO system, to estimate the BER.

\subsection{Equivalent SNR for SISO BER with phase noise}

Consider a unit-power QPSK modulated signal $s$ received with Gaussian phase noise $\varphi\sim \mathcal{N}(0,\sigma_\varphi^2)$ as 
\[
y(t) = e^{j\varphi(t)}s(t), \quad s \in \{\frac{1}{\sqrt{2}}(\pm1\pm j)\} .
\]
For Gray coded QPSK (which corresponds to independent BPSK streams along I and Q), the BER in the absence of additive noise is easily seen to be 
\[\text{BER} = Q(\frac{\pi/4}{\sigma_\varphi}). \]
This is the same BER produced by \textbf{additive} complex Gaussian noise with variance
\[
\sigma^2 = \frac{16}{\pi^2}\sigma_\varphi^2.
\]
We term this variance the ``equivalent \textit{additive} noise power" for \textit{phase} noise $\varphi$.
We now show that adding this variance to that of the additive noise yields an upper bound on BER.

\begin{theorem}
	\label{lemma:equivalent_noise}
	The BER of a Gray coded QPSK signal distorted by phase noise $\varphi\sim\mathcal{N}(0,\sigma_\varphi^2)$ and additive noise $n\sim \mathcal{CN}(0,\sigma_n^2)$,
	\[y = e^{j\varphi}s + n,\]
	is upper bounded by the BER achieved with \textbf{only additive} noise of variance
	\[
	\sigma^2 = \sigma_n^2 + \frac{16}{\pi^2}\sigma_\varphi^2,
	\]
	that is,
	\[\text{BER} \le Q\left(\frac{1}{\sqrt{\sigma_n^2 + \frac{16}{\pi^2}\sigma_\varphi^2}}\right).\] 
\end{theorem} 
\noindent\textit{Proof.} See Appendix \ref{append:lemma1proof}. 

\section{Multiuser Reception\label {sec:LMMSE}}

We employ LMMSE interference suppression followed by per-user constellation tracking, as described below.

\subsection{LMMSE reception}

In the absence of phase noise, the LMMSE receiver for the model (\ref{eq:rec_sig}) is given by
\begin{equation} \label{naive_LMMSE}
	\boldsymbol{\Gamma}_\text{LMMSE} = \mathbf{H}^H(\mathbf{H}\mathbf{H}^H + \sigma_\nu^2\mathbf{I})^{-1}.
\end{equation}
For the model (\ref{eq:reception_in_phase_noise_small_phase}) with phase noise, 
we derive the approximate LMMSE receiver by treating the channel dependent distortion, 
\[
\mathbf{z} = \left[
\begin{array}{c}
\mathbf{H}_1{j\phi_1}\\
\mathbf{H}_2{j\phi_2}\\
\vdots \\
\mathbf{H}_{N_T}{j\phi_{N_T}}\\
\end{array} \right]
\mathbf{s}
\]
as an additional noise term, where $\{ \phi_i \}$ are the independent VCO phase noise contributions at the output of tile PLLs. 

Assuming that the user symbols are unit power and uncorrelated, $\mathbb{E}\,\mathbf{s}\mathbf{s}^H = \mathbf{I}_K$ and the covariance matrix of this distortion is of the form
\[
\mathbf{C}_z = \sigma_\phi^2 \left[  
\begin{array}{c c c c}
\mathbf{H}_1\mathbf{H}_1^H & \mathbf{0} & \dots & \mathbf{0} \\
\mathbf{0} & \mathbf{H}_2\mathbf{H}_2^H & \dots & \mathbf{0} \\
\vdots & \vdots & \ddots & \vdots \\
\mathbf{0} & \mathbf{0} & \dots & \mathbf{H}_{N_T}\mathbf{H}_{N_T}^H 
\end{array}
\right]
\]
where $\sigma_\phi^2$ is the filtered VCO phase noise variance described by (\ref{eq:sigma_phi_2}).
The optimal linear receiver for a phase noise distorted system is therefore calculated as
\begin{empheq}[box=\widefbox]{align}
	\boldsymbol{\Gamma}_\text{LMMSE} = \mathbf{H}^H(\mathbf{H}\mathbf{H}^H + \sigma_\nu^2\mathbf{I} + \mathbf{C}_z)^{-1}.
	\label{eq:LMMSE_receiver}
\end{empheq}

In terms of implementation, the naive LMMSE receiver (\ref{naive_LMMSE}) may be realized using one-shot channel estimates at the beginning of a duration over which
the spatial channels are modeled as invariant, while the receiver (\ref{eq:LMMSE_receiver}) may be obtained by continuous adaptation over such an interval.

\begin{figure}
	\centering
	\includegraphics[width=0.6\columnwidth
	]{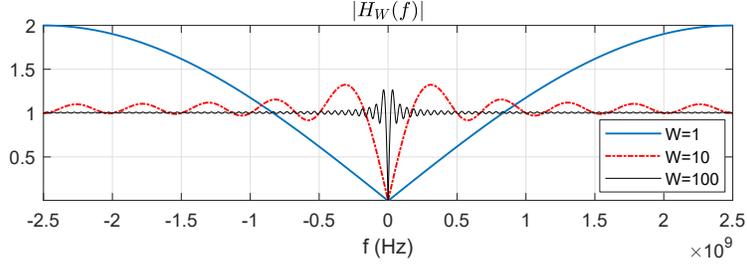}
	\caption{Frequency response of constellation tracking filter for different window sizes (with respect to phase signal). $W=1$ is equivalent to differential modulation.}
	\label{fig:derotation_filter}
\end{figure}

\begin{figure}
	\centering
	\includegraphics[width=0.6\columnwidth
	,trim={0.25in 0in 0.3in 0in},clip
	]{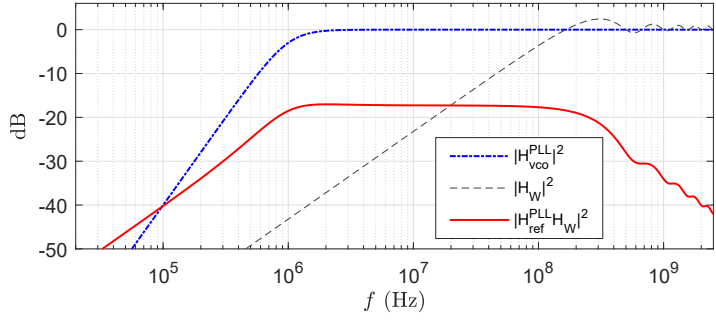}
	\caption{Phase noise spectrum shaping by PLL and constellation tracking filters.}
	\label{fig:reference_filtering}
\end{figure}


\subsection{Per-user constellation tracking}

Reference phase noise is common over the array and therefore passes through the linear receiver to the output of all channels:
\begin{equation*}
	{\mathbf{y}} = \boldsymbol{\Gamma}(\mathbf{x}e^{j\phi_0}) = e^{j\phi_0}\boldsymbol{\Gamma}\mathbf{x} .
\end{equation*}
Assuming the nearest neighbor estimates are correct, i.e., $\hat{s}_k = \text{NN}(y_k) = s_k$, output phase noise can be estimated after detection as \[\hat{\phi}_\text{out} = \angle\frac{y_k}{\hat{s}_k}.\] 
This overall phase distortion contains high pass contributions from tile VCOs and a lowpass component that is the filtered reference phase noise. By averaging the output phase noise estimate over several symbols, the low pass component is isolated and can be used to undo the reference phase drift for upcoming symbols. 
Assuming error-free symbol detection, this constellation tracking process can be abstracted as a windowed demeaning filter applied to the output phase noise, 
with impulse response

\[
h_W(t) = \delta(t) - \frac{1}{W}\sum_{i=1}^W \delta(t - i\, T_\text{symb})
\]
where $W$ is the demeaning window size and $T_\text{symb}=1/B
$ is the symbol duration. Window size is a tunable design parameter; a smaller window can track phase drift faster and has a larger rejection bandwidth, but introduces greater noise enhancement at high frequencies. This effect is clearly shown in Fig. \ref{fig:derotation_filter} where the frequency response of the tracking filter is depicted for different window sizes. High pass filtering by the constellation tracking mechanism compounded with the low pass filtering of the PLL diminishes almost all of the reference phase noise power. What remains can be formulated using the abstracted tracking model as

\begin{empheq}[box=\widefbox]{align}
	\sigma_0^2 = \int_{-B/2}^{B/2}L_\text{ref}(f)\left|H^\text{PLL}_\text{ref}(f)H_\text{W}(f)\right|^2df
	\label{eq:sigma_0_2}
\end{empheq}
which depends on the filtering bandwidths of the PLL and constellation tracking process. 
This value is reported in Table \ref{tab:q_coeffs} for the nominal configuration along with the total filtering coefficients applied to the $a_i$ components of $L_\text{ref}(f)$ satisfying $\sigma_0^2 = \sum_{i=0}^3 q_i a_i$. 

The impact of drift tracking on VCO phase noise is less significant as this is already a high pass signal. This impact is slight noise enhancement at the higher end of the spectrum (evident in Fig. \ref{fig:derotation_filter}) which, in our running example, increases output VCO phase noise by about $0.7$\%. If the bandwidth of $H_\text{vco}^{PLL}$ is smaller than that of $H_W$, drift tracking may \textit{decrease} the output VCO phase noise covariance.  It is worth noting that constellation tracking has no effect on the \textit{cross-user} interference caused by VCO phase noise since, (a) it is applied \textit{after} multiuser reception where crosstalk is produced, and (b) phase alteration does not alter the distribution of this interference as it is a zero-mean complex Gaussian variable.

\section{Interference Analysis \label{sec:interference}}
For the signal model (\ref{eq:reception_in_phase_noise_small_phase}), the output of a linear receiver
$\boldsymbol{\Gamma}$ is approximated by
\begin{align*}
	{\mathbf{y}} 
	& = \boldsymbol{\Gamma}\mathbf{H}\mathbf{s} + \boldsymbol{\Gamma}\boldsymbol{\nu} + \sum_{i=1}^{N_T}j\phi_i \boldsymbol{\Gamma}_i \mathbf{H}_i \mathbf{s}
\end{align*}
where $\boldsymbol{\Gamma}_i$ denotes the $i$'th $K\times N_0$ block of $\boldsymbol{\Gamma}$. That is,
\[\boldsymbol{\Gamma} = [\boldsymbol{\Gamma}_1\,\boldsymbol{\Gamma}_2\,...\,\boldsymbol{\Gamma}_{N_T}].
\]
The naive LMMSE receiver (\ref{naive_LMMSE}) focuses on suppression of the coherent (across tiles) interference in the first term. For our nominal load factor of $\beta = \frac{1}{4}$, and under the assumed minimum spatial frequency separation of $2 \pi/N$, the resulting LMMSE receiver does not lead to a significant attenuation of the desired signal.  The LMMSE receiver (\ref{eq:LMMSE_receiver}) does much the same, since the phase noise causing the third term is significantly smaller
(by design) than the coherent interference.  Furthermore, the third term is not coherent across tiles ($\{ \phi_i \}$ are independent), and tile-level interference suppression is not feasible in the scaling regime of interest to us, where the number of users is greater than the (fixed) number of antennas per tile.  We estimate tile-level interference,
therefore, under the following approximation.\\
{\bf Approximation:} {\it At the tile-level, both LMMSE variants are assumed to be aligned with the spatial matched filters for the users:}  
\begin{equation}
	\boldsymbol{\Gamma}_i \approx \frac{1}{N}\mathbf{H}_i^H.
	\label{eq:MF_approximation}
\end{equation}
We have verified this approximation via extensive simulations, but report only an example result: at SNR of 14 dB,
the normalized correlation at the tile-level between the LMMSE receiver and the spatial matched filter, at the tile level, is found to exceed 0.97 with probability 99\%.


Under the approximation (\ref{eq:MF_approximation}), we obtain (see Appendix \ref{appendix:correlations}) that the diagonal entries are equal to
\begin{equation} \label{diagonal}
	\left(\boldsymbol{\Gamma}_i \mathbf{H}_i\right)_{k,k} = \frac{N_0}{N}
\end{equation}
and the off-diagonal entries are zero-mean with variance 
\begin{align}
	\mathbb{E}\left|\left(\boldsymbol{\Gamma}_i \mathbf{H}_i\right)_{k,l}\right|^2 = \frac{N_0}{N^2},\qquad \qquad (k\neq l) \label{eq:offdiagonal}
\end{align}
We can now state the following result.

\begin{theorem}
	Under the approximation (\ref{eq:MF_approximation}), phase noise causes multiplicative self-noise and additive cross-user
	interference described by
	\begin{empheq}[box=\widefbox]{align}
		y_k \approx \gamma e^{j\psi}s_k + I
	\end{empheq}
	where 
	\begin{empheq}[box=\widefbox]{align}
		&\mathbb{E}\psi^2 = \frac{\sigma_\phi^2}{N_T}, \qquad \mathbb{E}\gamma = 1-\frac{1}{2}\sigma_\phi^2, \nonumber\\ 
		&\mathbb{E}|I|^2 = \frac{K-1}{N}\sigma_\phi^2 \approx \beta \sigma_\phi^2.
	\end{empheq}
	\label{theorem:phase_noise_interference}
\end{theorem}   
\noindent \textit{Proof.} 
Setting aside the additive contribution of thermal noise,
\begin{equation*}
	{\mathbf{y}} = \boldsymbol{\Gamma} \left[
	\begin{array}{c}
		\mathbf{H}_1e^{j\phi_1}\\
		\mathbf{H}_2e^{j\phi_2}\\
		\vdots \\
		\mathbf{H}_{N_T}e^{j\phi_{N_T}}\\
	\end{array} \right]
	\mathbf{s} .
\end{equation*}
The output of channel $k$ is a combination of contributions from the desired and interfering users which we separate as
\begin{equation*}
	y_k = \left( \sum_{i=1}^{N_T} (\boldsymbol{\Gamma}_i\mathbf{H}_i)_{k,k} e^{j\phi_i} \right) s_k + \sum_{l\neq k} \left(\sum_{i=1}^{N_T} (\boldsymbol{\Gamma}_i\mathbf{H}_i)_{k,l} e^{j\phi_i}\right) s_l.
\end{equation*} 
The multiplicative self-noise is derived using the second order Taylor expansions of sine and cosine functions:
\begin{align*}
	\sum_{i=1}^{N_T}(\boldsymbol{\Gamma}_i\mathbf{H}_i)_{k,k}e^{j\phi_i} &\approx \frac{N_0}{N}\sum_{i=1}^{N_T}\left(\cos \phi_i + j \sin \phi_i \right)  \approx \frac{1}{N_T} \sum_{i=1}^{N_T}(1- \frac{1}{2}\phi_i^2) + j \phi_i \\
	& \approx  \left(1- \frac{1}{2N_T}\sum_{i=1}^{N_T}{\phi_i^2}\right) + {j\left( \frac{1}{N_T} \sum_{i=1}^{N_T}\phi_i\right) }= \gamma+j\psi \approx \gamma e^{j\psi}
\end{align*}
where $\gamma = 1- \frac{1}{2N_T}\sum_{i=1}^{N_T}{\phi_i^2}$
is an average of identically distributed non-zero-mean variables, and is therefore well-approximated by its expected value,
\[
\gamma \approx \mathbb{E}\gamma = 1-\frac{\sigma_\phi^2}{2}
\]
and $\psi =\frac{1}{N_T} \sum_{i=1}^{N_T} \phi_i$ 
is an average of zero-mean i.i.d. Gaussian random variables, with variance
\[
\mathbb{E}\psi^2 = \frac{\sigma_\phi^2}{N_T}.
\] 

We evaluate cross-user interference by using the small-phase approximation (first order Taylor expansion) to obtain
\begin{align*}
	\sum_{i=1}^{N_T} (\boldsymbol{\Gamma}_i\mathbf{H}_i)_{k,l} \, e^{j\phi_i} & =  \sum_{i=1}^{N_T} (\boldsymbol{\Gamma}_i\mathbf{H}_i)_{k,l}  + \sum_{i=1}^{N_T}j\phi_i (\boldsymbol{\Gamma}_i\mathbf{H}_i)_{k,l} \\
	& = (\boldsymbol{\Gamma}\mathbf{H})_{k,l} +  \sum_{i=1}^{N_T}j\phi_i (\boldsymbol{\Gamma}_i\mathbf{H}_i)_{k,l} 
	\approx \sum_{i=1}^{N_T}j\phi_i (\boldsymbol{\Gamma}_i\mathbf{H}_i)_{k,l} 
\end{align*}
with $(\boldsymbol{\Gamma}\mathbf{H})_{k,l}\approx 0$ resulting from the LMMSE receiver effectively suppressing cross terms in $\boldsymbol{\Gamma}\mathbf{H}\mathbf{s}$ 
using all $N$ degrees of freedom.
The covariance of cross-user interference is thus found to be
\begin{align*}
	\mathbb{E}|I|^2 &= \mathbb{E}\left| \sum_{l\neq k} \left(\sum_{i=1}^{N_T} (\boldsymbol{\Gamma}_i\mathbf{H}_i)_{k,l} e^{j\phi_i}\right) s_l \right|^2 \\
	& = (K-1)\, N_T\, \sigma_\phi^2\, \mathbb{E}|(\boldsymbol{\Gamma}_i\mathbf{H}_i)_{k,l}|^2 \\
	& = (K-1)\, \frac{N_T N_0}{N^2}\,\sigma_\phi^2  = \frac{K-1}{N}\sigma_\phi^2 \lesssim \beta\sigma_\phi^2
\end{align*}
\qed

\begin{figure}
	\centering
	\includegraphics[width=0.6\columnwidth
	,trim={.3in 0 .3in 0},clip
	]{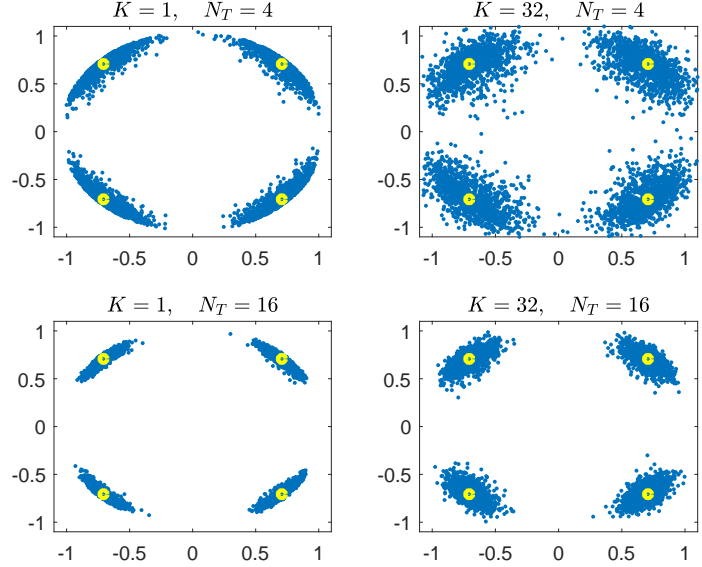}
	\caption{Scatter plot of received QPSK symbols on a $256$-element array for different load factors and number of tiles. The output phase noise decreases as number of tiles increases and interference is proportional to number of users. Additive noise has been set to zero to emphasize these effects. } \label{fig:scatter_plots}
\end{figure}

Accounting for all multiplicative and additive noise terms in the system, we arrive at the following model for the output signal:
\[{y}_k  = \gamma e^{j(\psi + \phi_0)}s_k + I + \nu'
\]	
where
\[\gamma = 1-\frac{\sigma_\phi^2}{2}, \qquad \mathbb{E}\psi^2 = \frac{\sigma_\phi^2}{N_T},  \qquad \mathbb{E}\phi_0^2 = \sigma_0^2,  \]
\[\mathbb{E}I^2 = \beta\sigma_\phi^2,  \qquad \, \mathbb{E}|\nu'|^2 = \frac{\sigma_\nu^2}{N}.\]
Fig. \ref{fig:scatter_plots} shows scatter plots of the received QPSK symbols in the I-Q plane for different values of $K$ and $N_T$.  Based on the preceding analysis, if load factor and tile size are kept constant, phase noise is not a bottleneck in scaling to larger arrays. The cross-user interference caused by phase noise only depends on the ratio of users to array size and is therefore constant, while output phase noise variance \textit{decreases} as the number of tiles grows. In fact, we expect performance to \textit{improve} as the system is scaled up, as long as loading, tile size, oscillator PSD, and beamformed SNR ($\sigma_\nu^2/N$) are fixed.
%

\vspace{2pt}\noindent\textbf{Predicting the BER of a phase-distorted system.} 
Using Theorem \ref{lemma:equivalent_noise}, we can determine the equivalent noise variance for our tiled system as
\begin{align}
	\sigma_\text{eq}^2 &= \frac{1}{(1-\frac{1}{2}\sigma_\phi^2)^2}\left(\frac{\sigma_\nu^2}{N} + \beta\sigma_\phi^2 + \frac{16}{\pi^2}\left(\frac{\sigma_\phi^2}{N_T} + \sigma_0^2 \right)\right)\nonumber 
\end{align}
This value is an expectation and actual SINR varies across users. Since we assume power leveling, this variation is primarily due to difference in cross-user interference.
In order to provide a pessimistic prediction, we substitute the average cross-talk power $\beta\sigma_\phi^2$ with $3$ standard deviations above its mean,
\[
\left(\beta + 3\, \frac{0.82 \sqrt{\beta}}{N_T}\right)\sigma_\phi^2
\]
(derivation of the variance of the cross-talk power omitted due to space considerations). Since BER is dominated by the worst-case,
we expect this pessimistic approach to be accurate.
The equivalent SINR is thus modified to
\begin{empheq}[box=\widefbox]{align}
	\sigma_\text{eq}^2 = \frac{1}{(1-\frac{1}{2}\sigma_\phi^2)^2} \Bigg(\frac{\sigma_\nu^2}{N} + (\beta + 2.46\frac{\sqrt{\beta}}{N_T})\sigma_\phi^2 
	+ \frac{16}{\pi^2}\left(\frac{\sigma_\phi^2}{N_T} + \sigma_0^2 \right)\Bigg). \label{eq:equivalent_variance} 
\end{empheq}

We now account for the reduction in SINR due to the reduction in signal power (or equivalently, noise enhancement) caused by suppression of the coherent interference.  At moderate load factors (e.g., $\beta = 1/4$) and SNRs, this SINR penalty is well approximated by that due to a zero-forcing receiver.  Let $\rho$ denote the normalized cross-correlation
between the spatial channels for two randomly chosen users.
If the spatial frequencies are uniform over $(-\pi,\pi)$, then $ E[ |\rho|^2]= \frac{1}{N}$ (see Appendix \ref{appendix:correlations}).  When we enforce a minimum spatial frequency separation
of $2 \pi/N$, we can actually show that $ E[ |\rho|^2] \leq  \frac{0.1}{N}$ (again, see Appendix \ref{appendix:correlations}).  In order to evaluate the impact of scaling, therefore, we set
\begin{equation} \label{rho}
	E[ |\rho|^2]= \frac{\alpha}{N}
\end{equation}
Now, denoting by $\rho_{lk}$ the normalized cross-correlation
between the spatial responses for users $l$ and $k$, the signal power $S$ relative to matched filtering, is bounded as follows:
\begin{equation} \label{signal_degradation}
	S \geq 1 -  \sum_{l\neq k} |\rho_{lk} |^2 \rightarrow 1 - (K-1) E[ |\rho|^2] \geq 1 - \alpha \beta
\end{equation}
where the limit is as $K, N \rightarrow \infty$ with $\beta$ fixed, and where we have used (\ref{rho}).  Using the right-hand side of (\ref{signal_degradation}) as an approximation,
we obtain the following pessimistic prediction of the
equivalent SINR 
\begin{empheq}[box=\widefbox]{align}
	\text{SINR}_\text{eq} \approx \frac{1-\alpha \beta}{\sigma_\text{eq}^2}. 
	\label{eq:equivalent_SINR}
\end{empheq}
which in turn yields a pessimistic approximation for BER \cite{poor1997probability}:
\begin{equation} \label{BER_estimate}
	\text{BER} \approx Q\left(\sqrt{\text{SINR}_\text{eq}}\right).
\end{equation}
This prediction, with $\alpha=0.1$ in (\ref{eq:equivalent_SINR}), is expected to be accurate in our regime of interest of $K > N_0$ and moderate SNR. In the underloaded regime, per-tile interference suppression becomes feasible and, at high enough SNR, partial per-tile interference suppression is incentivized even for $K > N_0$.  In such settings (not of interest in our scaling regime),
our prediction is still pessimistic, but is not as good of an approximation; i.e., performance can be significantly better than predicted.

In the next section we summarize our design framework, and provide simulation results to validate our analytical predictions.


\begin{figure}
	\centering
	\includegraphics[width=0.85\columnwidth
	,trim={.2in 0 .3in 0},clip
	]{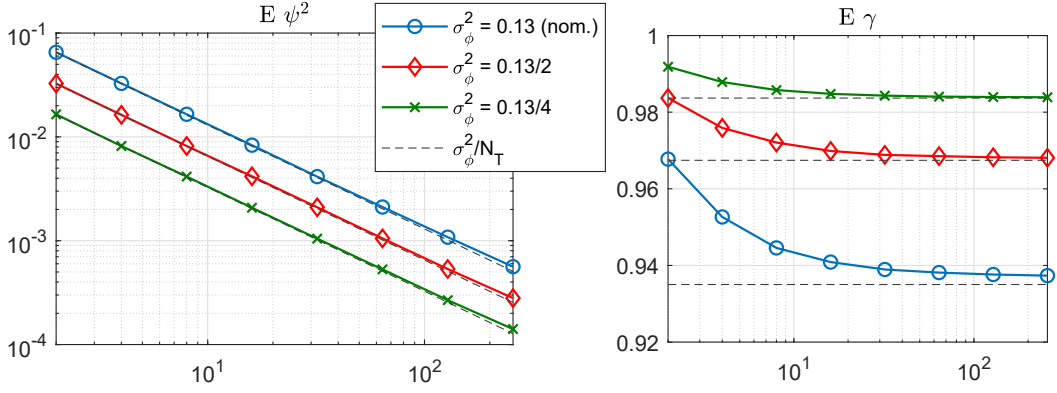}
	\caption{Self interference phase noise and amplitude attenuation for $256$ element array with 64 users.}
	\label{fig:cochannel}
\end{figure}

\begin{figure}
	\centering
	\includegraphics[width=0.65\columnwidth
	,trim={.19in 0 .3in 0},clip
	]{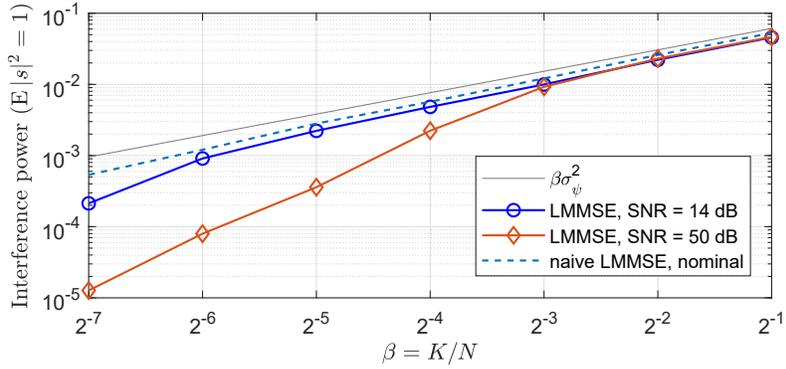}
	\caption{Scaling of cross-talk with load factor.  }
	\label{fig:crosschannel}
\end{figure}


\section{Numerical Results\label{sec:results}}

The models presented in this paper can be used to provide an \textbf{analytical cross-layer design framework} using four key observations.
\begin{itemize}
	\item The reference oscillator phase noise is filtered by the PLL and  constellation tracking filters and appears at the output of all channels with variance $\sigma_0^2 = \int L_\text{ref}(f)\left| H^\text{PLL}_\text{ref}(f) H_W(f)\right|^2df$. 
	
	\item VCO noise is filtered by the PLL resulting in tile phase noise with variance \[\sigma_\phi^2= \int L_\text{VCO}(f)\left| H^\text{PLL}_\text{vco}(f)\right|^2df.\] At the receiver output this produces phase noise of variance $\sigma_\phi^2/N_T$ and additive interference of power $\beta\sigma_\phi^2$.
	
	\item System performance is determined by parameters $\sigma_0^2$, $\sigma_\phi^2$, $N_T$, $\beta$, and $\sigma_\nu^2$ (beamformed SNR). A pessimistic prediction is given by the equivalent SINR
	\[\text{SINR}_\text{eq} = \frac{1-\alpha\beta}{\sigma_\text{eq}^2},\]
	\begin{align*}
		\sigma_\text{eq}^2 = (1-\sigma_\phi^2)^{-2}\Bigg(\text{SNR}^{-1} + (\beta + \frac{2.46\sqrt{\beta}}{N_T})\sigma_\phi^2
		+ \frac{16}{\pi^2}\left(\frac{\sigma_\phi^2}{N_T}+\sigma_0^2\right)\Bigg),
	\end{align*}
	where $\alpha=0.1$ for the specifications of our system model.
	
	\item {Permissible phase noise PSD can be expressed as a \textit{linear constraint} on the $L(f)$ coefficients, 
		\[\sum q_i a_i \le \sigma_\phi^2 \text{ or } \sigma_0^2, \]	
		with $q$ factors obtained by (\ref{eq:ph_component_linear}).}
\end{itemize}
Using the preceding guidelines, trade-offs between different design choices are modeled as simple analytical relationships that predict system performance with reasonable accuracy. 
In this section we provide numerical validation for the above framework and examine the scaling laws derived from it.

\begin{figure}
	\centering
	\includegraphics[width=0.6\columnwidth
	,trim={0.1in 0 .3in 0},clip
	]{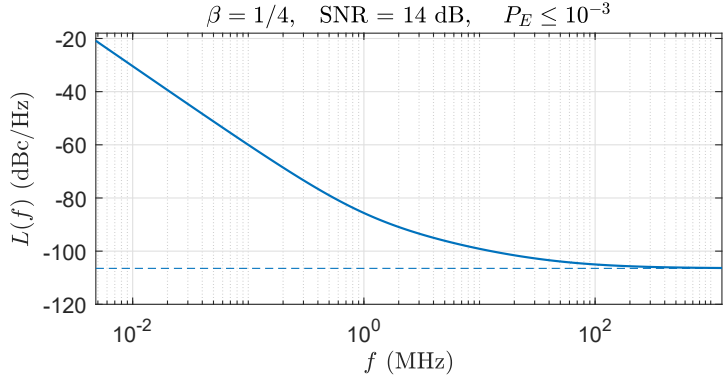}
	\caption{Acceptable phase noise mask for target BER $10^{-3}$ and $10$ Gbps rate in nominal system. Parameters: $a_0= 2.25\times 10^{-11}$ W/Hz, $a_1 = 9\times10^{-4}$ W, $a_2=9\times10^{2}$ WHz, $a_3=9\times10^{8}$ WHz$^2$. 
	}
	\label{fig:L_f_mask}
\end{figure}

We first 
provide an example of an acceptable \textbf{phase noise mask} for our nominal system specifications. We use the model of (\ref{eq:phase_noise_model}) to generate  phase noise signals.
The reference phase noise PSD is set lower than that of the VCO by a factor of $N_f$. 
In practice, the reference is likely to be a high quality crystal resonance oscillator with very low phase noise.
The shape of the curve ($a_i$ parameters) is chosen such that low-pass components have approximately the same combined impact as the constant component for our nominal system. 
Fig. \ref{fig:L_f_mask} shows the resulting $L(f)$ mask which indicates feasible phase noise requirements for our target system, as THz oscillators with lower $L(f)$ have been reported in the literature \cite{momeni2011high}. 

Using the PSD curve shown in Fig. \ref{fig:L_f_mask}, we isolate the two distortion terms described in Theorem \ref{theorem:phase_noise_interference} as follows. To measure \textbf{self-noise}, we set up the end-to-end multiuser system with $\beta=1/4$ and LMMSE reception (as described in Section \ref{sec:LMMSE}). We then set the transmitted symbol, $s_k(t)$, to zero for all but one user throughout each simulation sequence and set the additive noise to zero (but use the nominal value for $\sigma_\nu^2$ in deriving the LMMSE receiver). Fig. \ref{fig:cochannel} depicts self-noise 
for the designated user as a function of tile size and phase noise variance for a $256$-element array. 
The analytical predictions 
are also plotted for comparison. We observe that simulation results follow the analytical prediction closely. 

\begin{figure}
	\centering
	\includegraphics[width=0.48\columnwidth,trim={.3in 0 .4in 0},clip]{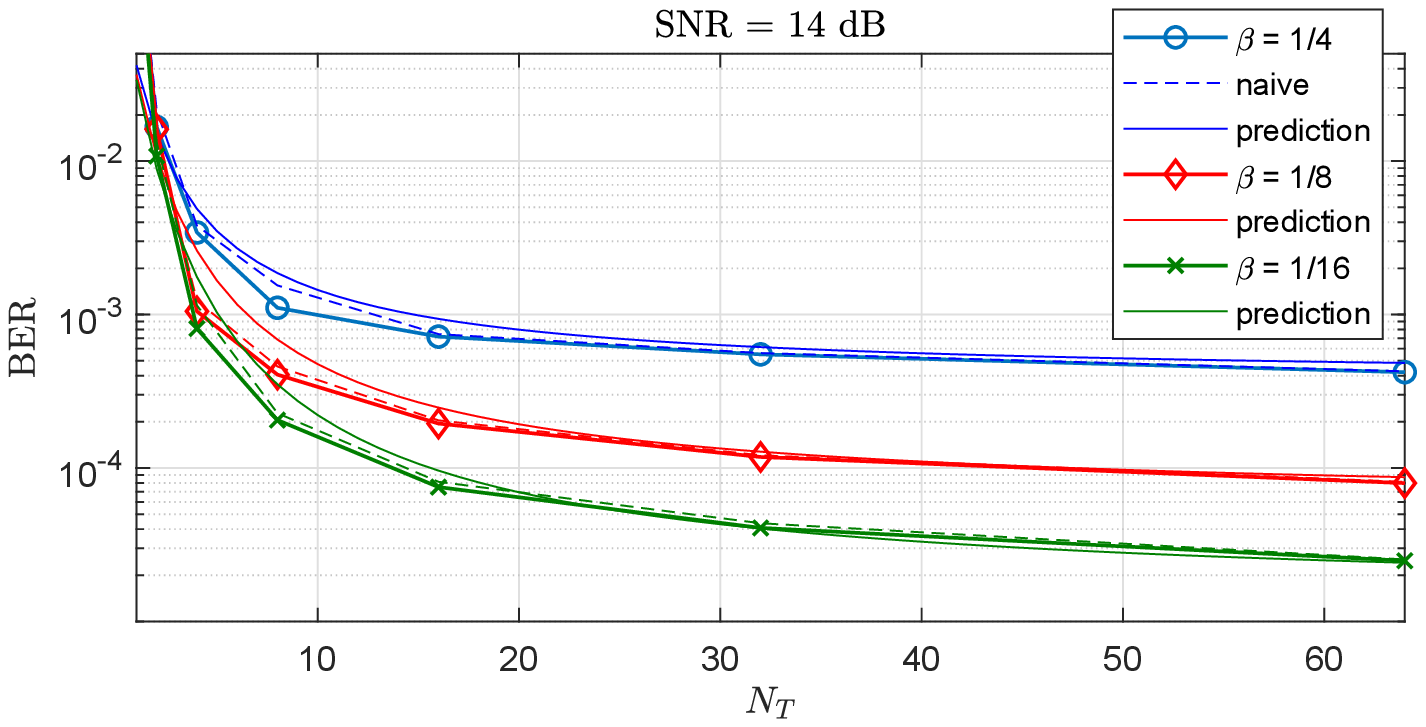}
	\includegraphics[width=0.48\columnwidth,trim={.3in 0 .4in 0},clip]{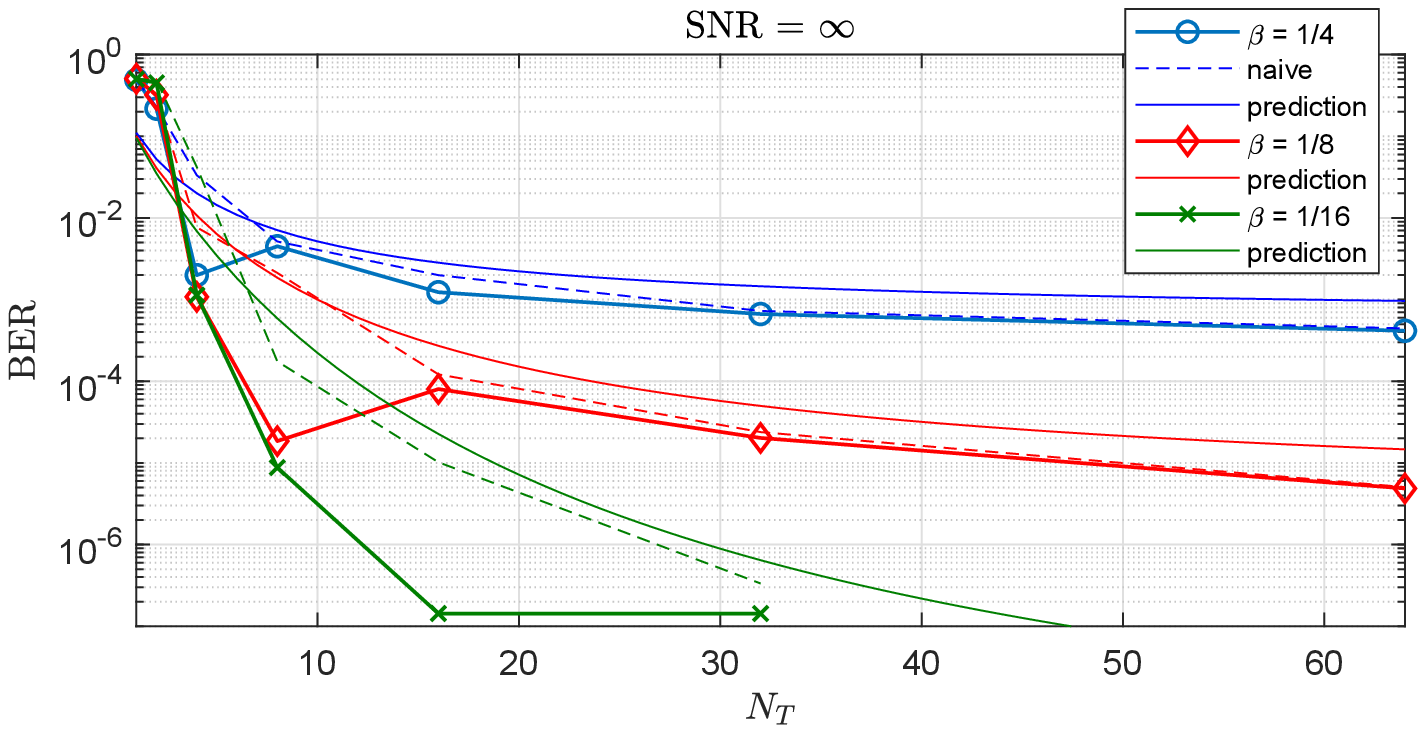}\\
	(a)$\qquad \qquad \qquad \qquad \qquad \qquad \qquad \qquad\qquad $(b)\\
	\caption{Full system simulation results for the nominal configuration using $L(f)$ function of Figure \ref{fig:L_f_mask} varying load factor and number of tiles ($N=256$ fixed). Optimal LMMSE, naive LMMSE (ignoring phase noise), and prediction of (\ref{eq:equivalent_variance}) and (\ref{eq:equivalent_SINR}) plotted for comparison.}
	\label{fig:full_system_sim}
\end{figure}

\begin{figure}
	\centering
	\includegraphics[width=0.6\columnwidth,trim={.1in 0 .3in 0},clip]{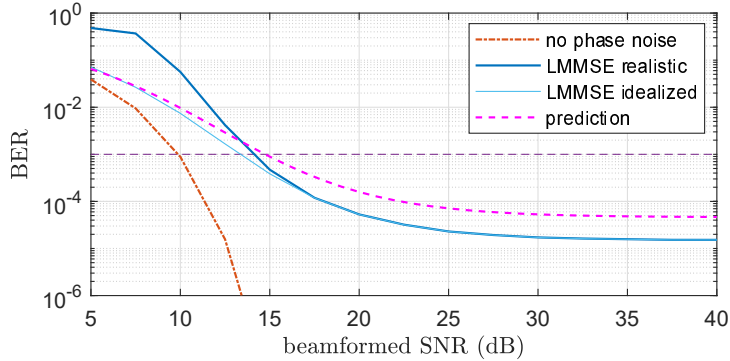}
	\caption{Performance (BER) as a function of SNR with and without phase noise. Solid narrow curve (light blue) assuming idealized constellation tracking, i.e., no error propagation. }
	\label{fig:scaling_with_snr}
\end{figure}


To isolate the \textbf{cross-user interference}, we perform a similar simulation but this time set the designated user's signal to zero while all other users remain active.
This way we only get cross-user interference signal at the designated output, the variance of which is the interference power. In Fig. \ref{fig:crosschannel} we report the cross-user interference variance averaged over many realizations for naive LMMSE (that ignores the effect of phase noise) and optimal LMMSE for nominal and high SNR. Both receivers follow the analytical prediction, $\mathbb{E}|I|^2=\beta\sigma_\phi^2$, fairly closely in most regimes. For an underloaded system at high SNR, analytical results {overestimate} interference for LMMSE reception since \textit{per tile} interference suppression becomes possible, as discussed in Section \ref{sec:interference}. 



\textbf{Full system simulations} that include the effects of VCO and reference phase noise, as well as realistic implementation of constellation tracking were performed to quantify the effect of system parameters on performance. 
Fig. \ref{fig:full_system_sim}a depicts performance as a function of load factor and tile size, assuming beamformed SNR of $14$ dB and phase noise PSD depicted in Fig. \ref{fig:L_f_mask}. When the array is divided into a larger number of tiles, self-noise is suppressed and BER decreases, but if $\beta$ remains constant the cross-user interference does not change with $N_T$ and therefore creates a performance floor. 
In Fig. \ref{fig:full_system_sim}b we depict the same results for a high SNR system. As expected, we see that BER is overestimated considerably for optimal LMMSE reception in the underloaded regime. 

Finally, Fig. \ref{fig:scaling_with_snr} shows how performance scales with beamformed link SNR in the presence and absence of phase noise. 
The performance floor caused by phase noise, especially its cross-user interference effect, is clearly visible in this figure. This floor can only be suppressed by reducing the load factor as shown in Fig. \ref{fig:full_system_sim}.
By comparing realistic drift tracking with the idealized model of (\ref{eq:sigma_0_2}) we see that, at low SINR, a decoding error can deteriorate tracking performance and cause additional errors in consecutive symbols. This effect can be mitigated by adding a margin to the target SINR. 
For our nominal setup, idealized performance is recovered with $0.5$ dB higher SINR, which is achieved by scaling down all noise terms uniformly, i.e., increasing SNR and reducing $L(f)$ each by $0.5$ dB. As expected, our analytical predictions are only pessimistic for the idealized case and may underestimate BER for a realistic system at low SINR. 

\section{Conclusions} \label{sec:conclusions}

Our analytical framework implies that, as far as the performance floor due to phase noise is concerned, a modular architecture with fixed-size tiles can be used to scale
all-digital mmWave multiuser MIMO up arbitrarily by increasing the number of tiles, keeping oscillator noise characteristics, bandwidth and load factor fixed. A naive LMMSE receiver that employs one-shot
channel estimates while ignoring phase noise, along with constellation tracking at the output, is found to work well.  Our analysis provides specific guidelines
to hardware designers on permissible phase noise PSD characteristics, and the specifications corresponding to our concept system are achievable in low-cost silicon processes.

While the scaling laws in Theorem \ref{theorem:phase_noise_interference} apply to arbitrary constellations
(e.g., see numerical results for 16QAM in our preliminary results \cite{rasekh2019phase}), the BER estimates based on
Theorem \ref{lemma:equivalent_noise} are specialized to Gray coded QPSK.  Extension of such estimates for larger constellations is an interesting
topic for future work.  While our analysis focuses on the impact of phase noise on demodulation, assuming ideal one-shot channel estimates are available, 
it is important to investigate how phase noise affects channel estimation. We expect reference phase noise to become more significant, since we can no longer count on attenuation due to post-demodulation constellation tracking. Such explorations are best undertaken in the context of specific receiver architectures. For example, the training period, and hence the impact of phase noise, can be significantly reduced by beamspace techniques which exploit the sparsity of the mmWave channel, as indicated in preliminary results reported in \cite{abdelghany2019beamspace}.  

The results here, along with analogous results regarding the impact of nonlinearities \cite{abdelghany2018towards}, indicate that hardware impairments such as phase noise and nonlinearities do not represent fundamental bottlenecks for scaling all-digital mmWave MIMO, and provide specific design guidelines for hardware front end design.  Of course, building RF hardware based on these design prescriptions is a significant challenge, as is management of the complexity of digital signal processing as bandwidth, number of antennas, and number of simultaneous users, scale up.  Again, beamspace techniques represent a promising framework \cite{abdelghany2019beamspace,abdelghany_globecom2019} for this purpose.

\section*{Acknowledgements}
This work was supported in part by ComSenTer, one of six centers in JUMP, a Semiconductor Research Corporation (SRC) program sponsored by DARPA. 

\appendices

\section{Proof of Theorem \ref{lemma:equivalent_noise}} \label{append:lemma1proof}
Gray coded QPSK corresponds to independent bits sent along I and Q. By symmetry, condition without loss of generality on transmitting $s=e^{j\pi/4}$, and on the probability of error in the I bit.
Consider the two signals
\begin{align*}
y_1 &= se^{j\varphi} + n \\
y_2 &= s + n_\varphi + n
\end{align*}
where $\varphi\sim\mathcal{N}(0,\sigma_\phi^2)$ and $n_\varphi\sim\mathcal{CN}(0,\frac{16}{\pi^2}\sigma_\phi^2)$, meaning real and imaginary parts of $n_\varphi$ are real zero-mean Gaussian variables with variance $\frac{8}{\pi^2}\sigma_\phi^2$. 

\begin{figure}
	\centering
	\includegraphics[width=0.7\columnwidth,trim={2in 2.5in 2.in 0.3in},clip]{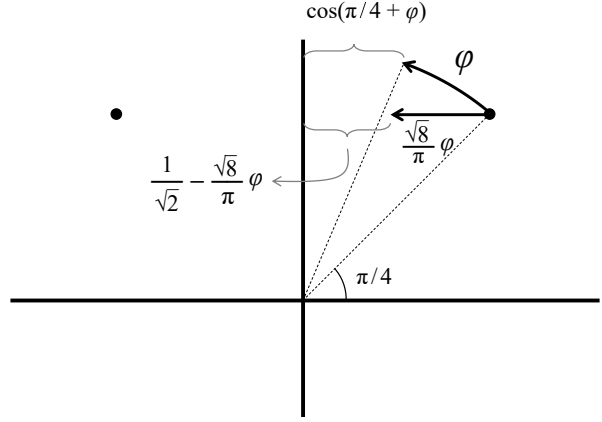}
	\caption{Margin left for additive noise $n$ after phase distortion and equivalent additive distortion.}
	\label{fig:equivalent_noise}
\end{figure}

An error in the I bit occurs when $y$ crosses the vertical boundary into the region $\mathfrak{Re}(y)<0$. 
In the first signal model, this error occurs when 
\[
\mathfrak{Re}(y_1) = \cos(\pi/4 + \varphi) + n_i <0
\]
and in the second model when
\[
\mathfrak{Re}(y_2) = \frac{1}{\sqrt{2}} + \mathfrak{Re}(n_\varphi) + n_i <0
\]
where $n_i$ is the real part of the complex Gaussian variable $n$. 
To compare the probability of these two occurrences we define the additive variable as
\[\mathfrak{Re}(n_\varphi) = -\frac{\sqrt{8}}{\pi}\varphi
\]
which is a zero-mean Gaussian that satisfies $\mathbb{E}\left(\mathfrak{Re}(n_\varphi)\right)^2 = \frac{8}{\pi^2}\sigma_\varphi^2$. The horizontal error margin left after the effect of $\varphi$ and $n_\varphi$ is, respectively, 
\[
m_1 = \cos(\pi/4 + \varphi) 
,\qquad 
m_2 = \frac{1}{\sqrt{2}} - \frac{\sqrt{8}}{\pi}\varphi,
\]
as depicted in Fig. \ref{fig:equivalent_noise}. For any $\varphi\in(0,\pi/4)$ we have $m_1-m_2\ge 0$ because $m_1-m_2$ is a convex function of $\varphi$ in this domain with value $0$ at the boundaries,
\begin{align*}
&@ \,\varphi=0 : \qquad m_1 = m_2 = 1/\sqrt{2}\\
&@ \,\varphi=\pi/4: \quad  m_1 = m_2 = 0 \\
& \frac{\partial^2}{\partial\varphi^2}(m_1 - m_2) = -\cos(\pi/4+\varphi) <0, \, \forall \varphi\in(0,\pi/4).
\end{align*}
We therefore conclude that
\[
\text{BER} = P[{\rm I~bit~wrong}] = \text{Pr}(\mathfrak{Re}(y_1)<0) \le \text{Pr}(\mathfrak{Re}(y_2)<0).
\]
\qed

\section{Computations of spatial inner products} \label{appendix:correlations}

Let $\ba ( \omega ) = (1, e^{j \omega},..., e^{j (n-1) \omega})^T$ denote the response of an $n$-element linear array to spatial frequency $\omega$. Note that
\begin{equation} \label{correlations1}
	|| \ba (\omega ) ||^2 =n
\end{equation}
Applying this for $n=N_0$ gives us (\ref{diagonal}).
The magnitude of the inner product between the responses for two different spatial frequencies $\omega_1$ and $\omega_2$, with $\Delta \omega = \omega_1 - \omega_2$, is given by
\begin{equation} \label{correlations2}
	\lvert \langle \ba ( \omega_1 ) ,   \ba ( \omega_2 ) \rangle \rvert = \lvert 1+e^{j \Delta \omega} + ...+e^{j (n-1) \Delta \omega} \rvert = 
	\left\lvert \frac{\sin n \Delta \omega/2}{ \sin \Delta \omega/2} \right\rvert
\end{equation}
The corresponding normalized inner product is the magnitude of the well-known Dirichlet kernel:
\begin{equation} \label{dirichlet}
	\kappa_n (\Delta \omega ) = \left\lvert \frac{\sin n \Delta \omega/2 }{n \sin \Delta \omega/2} \right\rvert
\end{equation}
If $\omega_1, \omega_2$ are independent and uniform over $(-\pi , \pi )$, then modulo $2 \pi$, $\Delta \omega$ is uniform over $(- \pi , \pi )$. 
Squaring the $| \cdot |$ term in (\ref{correlations2}) and taking expectations, the contribution of cross-terms of the form $e^{j k \Delta \omega}$, where $k$ is a nonzero integer, is zero.
We therefore obtain
\begin{equation} \label{correlations3}
	E \left[ |\langle \ba ( \omega_1 ) ,   \ba ( \omega_2 ) \rangle |^2 \right]= n
\end{equation}
Applying this for $n=N_0$ gives us (\ref{eq:offdiagonal}).
We may also write this as
\begin{equation} \label{correlations4}
	E\left[ |\kappa_n (\Delta \omega ) |^2 \right] = \frac{1}{n}
\end{equation}
Applying this for $n=N$ gives us $a=1$ in (\ref{rho}).

To see what happens when we enforce a minimum spatial separation $2 \pi /n$, note
that $|x| \geq |\sin x|$, so that $\kappa_n (x)  \leq | \frac{\sin (nx/2)}{nx/2}|$. We can now calculate that
\begin{align*}
	\frac{1}{2 \pi} \int_{-2\pi/n}^{2\pi/n} & \kappa_n^2 (x)  dx \ge 
	\frac{1}{2\pi}\int_{-2\pi/n}^{2\pi/n} \left|\frac{\sin(nx/2)}{nx/2}\right|^2 dx \\ & = \frac{1}{n\pi}\int_{-\pi}^{\pi} \left|\frac{\sin(u)}{u}\right|^2 du \ge \frac{0.90}{n}.
\end{align*}
Setting $n=N$, under the assumed minimum spatial separation of $2\pi/N$, we therefore obtain that
\[
\mathbb{E}|\rho|^2 \le \frac{0.1}{N}
\] 
so that we may set $\alpha=0.1$ in (\ref{rho}).

\bibliographystyle{IEEEtran}
\bibliography{PhaseNoiseJournal_singleColumn}

\end{document}